\documentclass[aps,prd,groupedaddress,showpacs,showkeys,amsmath,amssymb]{revtex4}
\usepackage{verbatim,graphicx,amssymb,amsbsy,bm,amsmath,rotating,epsfig}
\usepackage{psfrag}
\usepackage{hyperref}
\newcommand{\be}{\begin{equation}}
\newcommand{\ee}{\end{equation}}
\newcommand{\bea}{\begin{eqnarray}}
\newcommand{\eea}{\end{eqnarray}}

\begin{document}
\title{Warm dark matter sterile neutrinos in electron capture and beta decay spectra}
\author{O. Moreno}
\affiliation{Departamento de F\'{\i}sica At\'omica, Molecular y Nuclear, Facultad de Ciencias F\'{\i}sicas, Universidad Complutense, 28040 Madrid, Spain}
\author{E. Moya de Guerra}
\affiliation{Departamento de F\'{\i}sica At\'omica, Molecular y Nuclear, Facultad de Ciencias F\'{\i}sicas, Universidad Complutense, 28040 Madrid, Spain}
\author{M. Ram\'on Medrano}
\affiliation{Departamento de F\'{\i}sica Te\'orica I, Facultad de Ciencias F\'{\i}sicas, Universidad Complutense, 28040 Madrid, Spain}
\date{\today}

\begin{abstract}
We briefly review the motivation to search for sterile neutrinos in the keV mass scale, as dark matter candidates, and the prospects to find them in beta decay or electron capture spectra, with a global perspective. We describe the fundamentals of the neutrino flavor-mass eigenstate mismatch that opens the possibility of detecting sterile neutrinos in such ordinary nuclear processes. Results are shown and discussed for the effect of heavy neutrino emission in electron capture in Holmium 163 and in two isotopes of Lead, 202 and 205, as well as in the beta decay of Tritium. We study the de-excitation spectrum in the considered cases of electron capture and the charged lepton spectrum in the case of Tritium beta decay.
For each of these cases, we define ratios of integrated transition rates over different regions of the spectrum under study,  and give new results that may guide and facilitate the analysis of possible future measurements, paying particular attention to forbidden transitions in Lead isotopes.

\end{abstract}

\maketitle

\section{Introduction \label{introduction}}
There is an inconsistency between the amount of matter inferred from gravitational effects and the one we see on observable scales. This fact leads to consider the existence of dark matter (DM). Evidence arises from astrophysical and cosmological probes \cite{zwi33,cosmo}, such as the kinematics of virially bound systems, rotation curves of spiral galaxies, strong and weak lensing, cosmic microwave background (CMB) information on matter density and geometry of the universe, mass-to-light ratios in dwarf spheroidal galaxies (dSph), and large surveys to measure universe structures. Because of the limits inferred from Big Bang nucleosynthesis, an important fraction of DM should be non baryonic. Although the nature of DM is still unknown, a popular hypothesis considers that DM consists of elementary particles \cite{cosmo}. DM is the main component of galaxies, being in average at least six times more abundant than baryonic matter (more than 81$\%$ of the matter in the Universe), but its nature is still unknown. DM self-interactions have been unobserved so far, and DM particle candidates are bound by gravitational interactions. Very distinct predictions for small scale structures in the Universe (below 100 kpc) are obtained by DM particles with different mass scales \cite{les12}. On the large mass scale side, WIMPs (weakly interacting massive particles) are popular candidates to DM that have masses of the order of GeV or even TeV. This type of particles fall in the category of cold dark matter (CDM), which predicts, for small scales, too many galaxy satellites in the Milky Way and cusped profiles for the mass density of galaxies, contradicting present observational evidences. On the contrary, particles with mass in the keV scale, namely warm dark matter (WDM) \cite{boy08}, are able to reproduce the number of observed satellite galaxies, as well as the cored profiles found in DM-dominated objects such as dwarf spheroidal galaxies.

Concerning the satellite problem, cosmic structures would form from the gravitational collapse of overdense regions in the DM primordial field. Free relativistic particles do not cluster, and structures at scales smaller than the particle free-streaming length $l_{fs}$ are erased. The free-streaming length is the distance travelled freely by a relativistic particle after decoupling from the primordial plasma due to Universe expansion (approximately the distance travelled before the transition to non-relativistic velocities). WDM particles (keV scale) give $l_{fs}\sim$ 100 kpc, while CDM (GeV-TeV scale), which are heavier and slower than WDM, would give a $l_{fs}$ a million times smaller, and lead to the existence of a host of small scale structures \cite{kolb90}. On the galaxy density profiles ($\rho$), CDM gives a steep cusp at the center ($\rho \sim r^{-1}$) \cite{cusped}.  On the contrary, WDM gives a finite constant density core at the center ($\rho\sim\rho_0$), in agreement with observations \cite{cored}.  WDM quantum effects \cite{des13} could be important inside the  galaxy core (below 100 pc), showing a fermionic nature for the DM particle through the manifestation of quantum non vanishing pressure versus gravity.

Astrophysical observations from DM-dominated objects, as well as theoretical analysis, lead to a DM fermionic thermal particle with a mass around 2 keV \cite{men16}. Chandra and XMM-Newton detections in the X-ray spectra of the M31 galaxy and the Perseus cluster (both DM-dominated) seem to be consistent with sterile neutrinos of a few keV. In particular, an unidentified 3.55 keV line has been observed that seems compatible with the decay or annihilation of sterile relic neutrinos \cite{bulboy14} and that does not correspond to any known atomic emission. Although the interpretation of this line is still subject to debate, it could be considered an indication of the decay of a 7.1 keV non-thermal sterile neutrino \cite{aba14}.

As it is well known, the Standard Model (SM) of elementary particles does not describe DM particles, nor does it provide a mass for active neutrinos ($\nu_e$, $\nu_{\mu}$, $\nu_{\tau}$). In this model, neutrino eigenstates are left-handed, and they transform as doublets under the weak SU(2) gauge group. In extended SM models \cite{dol02}, additional SU(3) x SU(2) x U(1) singlet right-handed neutrinos are introduced. Neutrino mass eigenstates will be linear combinations of the left and right-handed states, and mass matrix eigenvalues will split into lighter and heavier states (seesaw mechanism). The lighter mass eigenstates would be the main components of the active flavor eigenstates of the neutrinos, whereas the heavier ones would be dominant in sterile neutrino flavors \cite{pon68}. Some extended SM models add three extra sterile neutrinos, one with a mass of the order of keV (see \cite{mer13} and references therein). Neutrinos in the keV mass scale open the possibility to detect warm dark matter in nuclear electron capture and beta decay. Heavier sterile neutrinos cannot be produced in weak nuclear decays, and there are of course many other candidate particles for DM unconnected in principle to weak nuclear processes, such as the lightest supersymmetric particles \cite{lsp}.

In this paper, we assume that a keV neutrino, a sound candidate for DM \cite{adh16}, is produced in ordinary weak nuclear processes, such as beta decay and electron capture, via neutrino mixing.
Experiments searching for active neutrino masses, also look for sterile neutrinos in the keV range \cite{drex13}: MARE \cite{mar} (that used Rhenium 187 beta decay and is not active anymore); KATRIN \cite{kat}, PTOLEMY \cite{ptolemy} and Project8 \cite{project8} (using Tritium beta decay); and ECHo \cite{echo} and HOLMES \cite{holmes} (using Holmium 163 electron capture). 
The mass of the neutrino (or antineutrino) emitted in a weak nuclear decay has an effect on the energy spectrum of the process, as was first shown by Enrico Fermi in the early 1930s \cite{fer34}. For the active neutrinos this effect shows up at the endpoint of the spectrum, whereas for the sterile neutrinos considered here it may be expected to appear at a few keV below the endpoint. Clearly, in order to leave a fingerprint the sterile neutrino mass must be within the $Q$ window, {\it i.e.}, lower than the energy $Q$ available in the decay. The energy spectrum to be analyzed corresponds to the emitted charged lepton in the case of beta decay, namely, the spontaneous conversion of a neutron into a proton or viceversa with emission of the charged lepton and an antineutrino or neutrino. In the case of electron capture only a neutrino is emitted and no charged lepton comes out, so that the spectrum to be measured corresponds to the de-excitation of the daughter atom.

A summary of current experimental studies of neutrino properties in the frontiers of intensities and sensitivities is given in \cite{kos15}. Experiments are under way to determine directly the active neutrino mass from neutrinoless double beta decay \cite{giu12} as well as from single beta decay \cite{kat,ptolemy} and electron capture \cite{echo,holmes}. The last two types of experiments also provide a way to search for WDM sterile neutrinos in the measured spectra. In electron capture experiments, the spectrum collected in a calorimeter is directly linked to the excitation energies of the daughter atoms or molecules. In a calorimeter the active source is embedded in the detector, which collects the energy of all the particles emitted in the de-excitation processes that take place in the source, except that of the neutrinos. In beta decay, the fact that the atoms or molecules may remain excited poses a challenge on the interpretation of the electron spectrum. Electron capture experiments take advantage of the larger statistics in the spectrum regions around the capture resonances. A limitation of electron capture measurements is that in a calorimeter, where the full energy range of the spectrum is measured, pile-up can be a problem that should be prevented by limiting the activity in the experimental setup.
We refer always in this paper to Earth-based experiments where both the emission and the detection of particles take place in the laboratory. Other possible scenarios, like the search for sterile neutrinos in stellar matter (stellar beta decay and electron capture rates), are beyond the scope of this paper. The phase space for these rates increases manifold in stellar environment. For a reference on stellar weak rates, see \cite{stellar}.

\section{Heavy mass and sterile flavor in neutrino states \label{mixing}}

Already observed neutrino oscillations (see for instance Ref. \cite{osci}) among light active neutrinos are due to the fact that the neutrino flavor eigenstates and the mass eigenstates are not the same. Each flavor eigenstate, associated to a charged lepton, can be written as a combination of mass eigenstates, and vice versa \cite{pet13}. For instance, the neutrino flavor eigenstate emitted after electron capture, called electron neutrino ${\nu}_{e}$, can be written as a combination of the three light SM mass eigenstates and, hypothetically, of one or more extra, heavier mass eigenstates as \cite{dk,shr}
\begin{equation}
|\nu_{e}\rangle = U_{e1} \:|\nu_{1}\rangle + U_{e2} \:|\nu_{2}\rangle + U_{e3} \:|\nu_{3}\rangle +\sum_{h} U_{eh} \:|\nu_{h}\rangle
 \label{mixing}   
\end{equation}
where the subscript $h=\{4,5,...\}$ stands for extra (heavier) mass eigenstates, and the quantities $U$ belong to the unitary neutrino mixing matrix. The masses of the three light SM mass eigenstates are so close to each other that so far no measurement has been able to discern which of them has been emitted in a given process. As a result, the three are emitted as a coherent superposition, which is at the origin of the neutrino oscillation phenomenology. The phase of each mass eigenstate changes at a different rate while travelling, giving rise to a different superposition at each location that translates into a varying (oscillatory) probability of detection of a given flavor eigenstate.

To simplify things we will consider here the linear combination of light mass eigenstates as a single, effective neutrino mass eigenstate, called `light', with mass
\begin{equation}
m_l = \overline{m}_{\nu_{e}} = \left( U^2_{e1} \:m^2_{1} + U^2_{e2} \:m^2_{2} + U^2_{e3} \:m^2_{3} \right)^{1/2} \;,
\label{effective_mass}   
\end{equation}
and just one extra mass eigenstate, clearly heavier than the others, called `heavy' (with mass $m_h$, that we shall consider in the keV range):
\begin{equation}
|\nu_{e}\rangle = \cos\zeta \:|\nu_{l}\rangle + \sin\zeta \:|\nu_{h}\rangle \:,
 \label{nu_elec_composition}   
\end{equation}
where the mixing amplitudes have been written in terms of a mixing angle $\zeta$ between the light and the heavy neutrino mass eigenstate. The other possible combination of these two mass eigenstates would be the sterile flavor eigenstate:
\begin{equation}
|\nu_{s}\rangle = -\sin\zeta \:|\nu_{l}\rangle + \cos\zeta \:|\nu_{h}\rangle \:,
 \label{nu_sterile_composition}   
\end{equation}
such that $\langle \nu_{e} |\nu_{s} \rangle=0$. For the sterile neutrinos that can be relevant as WDM, cosmological constraints based on the observed average dark matter density suggest that the value of the mixing angle could approximately be $\zeta=0.006^{\circ}$ \cite{dk}, corresponding to a flavor-mass amplitude $U_{eh} = \sin\zeta \approx 10^{-4}$. Other recent cosmological constraints give values of $\sin^2 (2\zeta)$ between 2 $\cdot$ $10^{-11}$ and 2 $\cdot$ $10^{-10}$ \cite{bulboy14}.

Sterile neutrino states with masses close to the ones of the active neutrinos can also have an impact on the patterns measured in oscillation experiments \cite{con13}. Heavier sterile neutrinos, as in particular in the keV scale, do not modify the oscillation patterns since they are not emitted coherently with the active neutrinos due to the large mass splitting.

The differential energy spectrum of a weak process where an electronic neutrino (or antineutrino) is emitted can therefore be decomposed in a term for light neutrino emission and another term for heavy neutrino emission as follows \cite{shr}:
 \begin{equation}
\frac{d\lambda}{dE} =  \:\frac{d\lambda^{l}}{dE} \; \cos^{2}\zeta
+ \:\frac{d\lambda^{h}}{dE} \; \sin^{2}\zeta
 \label{spectrum_mixing}   
\end{equation} 
This energy spectrum can be the one of the electron emitted in beta decay or the one of the daughter atom de-excitation in electron capture. The heavy mass eigenstate, if it exists, would be emitted independently of the other masses (non coherently) as long as the energy resolution of the detector is better than the mass difference, $\Delta \epsilon < (m_h-\overline{m}_e)$. Its contribution to the measured spectrum, $d\lambda^{h}/dE$, would show up in the range where the collected energy is small enough for the heavy neutrino mass to have been produced, namely when $0 \le E \le (Q-m_h)$, where $Q$ is the difference between the masses of the initial and the final atoms when the reaction takes place in vacuum. At the edge of that region, at $E=Q-m_h$, a kink in the spectrum would be found with the size of the heavy neutrino contribution, namely proportional to $\sin^{2}\zeta\sim \zeta^{2}$ (the latter approximation valid for small mixing angles, as is the case in realistic scenarios). 

\section{Theory of electron capture \label{capture}}

Let us consider the capture of an atomic electron by the nucleus $X$ ($Z$, $N$) to turn into the nucleus $Y$ ($Z-1$, $N+1$). The reactions involving the corresponding atoms (represented by the symbol of the nucleus within brackets) begin with electron capture, 
\begin{eqnarray}
(X) \:\rightarrow \:(Y)^{H}  + \nu_{i}  \;,
\label{atomic_capture}
\end{eqnarray}
followed by de-excitation of the daughter atom. In Eq. (\ref{atomic_capture}), $\nu_{i}$ is a neutrino mass eigenstate ($\nu_l$ or $\nu_h$) and the superscript $H$ accounts for the
excited state of the atom $(Y)$ corresponding to an electron hole in the shell $H$, due to the electron capture from this shell in the parent atom. The de-excitation of the daughter atom after electron capture can happen either through emission of a photon (X-ray emission):
\begin{eqnarray}
(Y)^{H} \:\rightarrow \:(Y) + \gamma \:,
\label{em_deexc}
\end{eqnarray}
or through emission of electrons (Auger or Coster-Kronig process) and photons:
\begin{eqnarray}
&& (Y)^{H} \:\rightarrow \:(Y^+)^{H',H''} + e^- \:\rightarrow \:(Y^+) + \gamma + e^- \:,\label{auger_deexc}  
\end{eqnarray}
where $H'$ and $H''$ represent holes in the electron shells of the ion $(Y^+)$.

The energies carried by the emitted photons and electrons are deposited in the calorimeter that surrounds the source and that measures the full energy spectrum of these particles, provided that the corresponding de-excitation lifetimes are much smaller than the detector time response \cite{nuc15}. The process in Eqs. (\ref{em_deexc}) and Eq. (\ref{auger_deexc}) yield the same calorimeter spectrum, {\it i.e.}, the peaks appear at the same energy values corresponding to the excitation energies, $E_H$, of the excited daughter atom $(Y)^{H}$. The excitation energy is the difference between the binding energies of the captured electron shell and the additional electron in the outermost shell. The former refers to the daughter atom, whereas the latter refers to the parent: $E_H\approx |B_H^{(Y)}|-|B_{out}^{(X)}|$. The reason is that for the shells above the vacancy the effective charge is closer to the one in the parent atom (there is one proton less in the nucleus but also one electron less in an inner shell) and therefore its binding energies should be used \cite{rob15}.

An alternative capture process to the one in Eq. (\ref{atomic_capture}) involves the instantaneous formation of a second hole $H'$ due to the mismatch between the spectator electron wave functions in the parent and in the daughter atom. Whereas the hole $H$ is left by the captured electron and therefore fulfils the energy and angular momentum conditions for such process, the extra hole $H'$ has a different origin, namely the above mentioned incomplete overlap of electron wave functions, which may occur in many electron shells. The electron leaving the extra hole may have been `shaken up' (excited) to an unoccupied orbital in the daughter atom:
\begin{eqnarray}
(X) \:\rightarrow \:(Y)^{H,H'} + \nu_{i}  \;,
\label{shakeup}
\end{eqnarray}
or it may have been `shaken off' to the continuum (ejected):
\begin{eqnarray}
(X) \:\rightarrow \:(Y^+)^{H,H'}  + e^- + \nu_{i}
\label{shakeoff}
\end{eqnarray}
In a shake-up process the de-excitation of the final atom contributes to the calorimeter energy with a peak at an energy approximately equal to the sum of the binding energies of the $H$ and $H'$ electrons, $E_{H,H'}\approx |B^{(Y)}_{H}|+|B^{(X)}_{H'}|-|B^{(X)}_{out}|$. Thus, satellite peaks show up located after the single-hole peak at $E_H$. In the case of a shake-off the calorimeter energy is the combination of $E_{H,H'}$ and the electron kinetic energy, the latter showing a continuous distribution as corresponds to the three-body decay of Eq. (\ref{shakeoff}).

It is important to notice that some electron emission processes of the type:
\begin{eqnarray}
(X) \:\rightarrow \:(Y^+)^{H',H''}  + e^- + \nu_{i}  \;,
\label{instant_electron2}
\end{eqnarray}
where none of the holes in the daughter atom correspond to the captured electron $H$, are quantum mechanically indistinguishable \cite{ruj13} from the three-step process in Eqs. (\ref{atomic_capture}) and (\ref{auger_deexc}), and therefore they contribute to the same one-hole peaks at $E_H$.
The contributions to the spectrum of the two-hole excitations that are actually distinct from the one-hole excitations (Eqs. (\ref{shakeup}) and (\ref{shakeoff})) have been a subject of recent studies devoted to search for the electron neutrino mass, and are therefore focused on the endpoint of the spectrum (see for instance \cite{rob15,ruj13,fae16}). For heavy neutrino searches a similar analysis should be performed at other regions of the energy spectrum, not just at the endpoint. In this paper we restrict the calculations to the one-hole states.

The density of final states of the emitted neutrino in electron capture is proportional to:
\begin{equation}
\rho = \rho_{\nu}(E_{\nu})  \propto \:p_{\nu} \:E_{\nu} = (E_{\nu}^2-m_{\nu}^2)^{1/2} \:E_{\nu} \:,
\end{equation}
where $E_{\nu}$, $p_{\nu}$, $m_{\nu}$ are the neutrino total energy, momentum and mass, respectively. On the other hand, the probability of capture of a bound electron follows a Breit-Wigner distribution of width $\Gamma_x$ and peaks at the energy $E_x$:
\begin{equation}
P(E) = \frac{\Gamma_x/2\pi}{(E-E_x)^2+\Gamma_x^2/4} \:,
\end{equation}

Using the Fermi's Golden Rule, the differential reaction rate with respect to the neutrino energy yields:
\begin{equation}
\frac{d\lambda}{dE_{\nu}} = K_{EC} \:(E_{\nu}^2-m_{\nu}^2)^{1/2} \:E_{\nu} \:\sum_H W^{(\nu)}_{H} \:\frac{\Gamma_H/2\pi}{[E_{\nu}-(Q-E_H)]^2+\Gamma_H^2/4}
\label{diff_decay_rate1}
\end{equation}
where $K_{EC}$ contains, among other factors, the weak interaction coupling constant and the nuclear matrix element. The factor $W_H^{(\nu)}$ is the squared leptonic matrix element for a given hole state $H$.

We write $W_H^{(\nu)}=C_H\:S^{(\nu)}_{H}$ to have a general expression for allowed and forbidden transitions. The factor $C_H$ is the squared amplitude of the bound-state electron radial wave function at the nuclear interior, containing also the squared overlap between the initial and the final atom orbital wave functions and the effect of electron exchange (as defined in Appendix F-2 of \cite{toi}). The cases discussed in the next sections involve allowed Gamow-Teller transitions ($^{163}$Ho (7/2$^-$) $\to$ $^{163}$Dy (5/2$^-$)) and first forbidden Gamow-Teller transitions ($^{202}$Pb (0$^+$) $\to$ $^{202}$Tl (2$^-$) and $^{205}$Pb (5/2$^-$) $\to$ $^{205}$Tl (1/2$^+$)). In the first case the factor $S^{(\nu)}_{H}=$ 1 for all $H$ states. In the second case the factor $S^{(\nu)}_{H}=$ 1 for electrons captured with orbital angular momentum $l=1$ (and neutrinos emitted with $l=0$) and $S^{(\nu)}_{H}=p_{\nu}^2$ for electrons captured with orbital angular momentum $l=0$ (and neutrinos emitted with $l=1$). The reason for this is that, due to conservation of total angular momentum, first forbidden transitions have $L=1$ for the leptonic pair. Hence, there is a linear momentum dependence in the leptonic matrix element corresponding to the lepton that carries the angular momentum in each specific capture. The corresponding $p_{e}^2$ factor for the $l=1$ electrons is embedded in the $C_H$ factor \cite{toi}.

Eq. (\ref{diff_decay_rate1}) includes every possible orbital electron capture, the probability of each peaking at $E_{\nu}=Q-E_H$, as dictated by energy conservation, where $E_H$ is the excitation energy of the final atom due to the electron hole $H$ resulting from capture.
The energy collected by a calorimeter from the atomic de-excitations is $E_c=Q-E_{\nu}$, namely all the available energy except for the one carried away by the neutrino. The differential rate can be written in terms of the calorimeter energy as:
\begin{equation}
\frac{d\lambda}{dE_c} = K_{EC} \:[(Q-E_c)^2-m_{\nu}^2]^{1/2} \:(Q-E_c) \:\sum_H \:C_{H}\:S^{(\nu)}_{H} \:\frac{\Gamma_H/2\pi}{(E_c-E_H)^2+\Gamma_H^2/4}
\label{diff_decay_rate2}
\end{equation}

In the next section we shall also consider integrated rates over particular energy ranges, which can be calculated numerically from Eq. (\ref{diff_decay_rate2}). A convenient compact analytical expression of the integrated rate can be obtained from Eq. (\ref{diff_decay_rate2}) by replacing the Breit-Wigner distributions by Dirac deltas. With this approximation, the integral of the differential rate can be written as
\begin{equation}
\lambda = K_{EC} \:\sum_H \:C_{H}\:S^{(\nu)}_{H} \:[(Q-E_H)^2-m_{\nu}^2]^{1/2} \:(Q-E_H)
\end{equation}
In this theoretical model we have neglected two-hole peaks, de-excitations through virtual intermediate states and interferences between de-excitation channels. The theoretical calorimeter spectrum is thus a single-hole approximation that assumes full collection of de-excitation energy by the calorimeter and no pile-up.

\section{Sterile neutrino effect in the spectrum \label{sterile_effect}}

Due to the smallness of the mixing angle it is useful to consider ratios between rates that emphasize the contribution of the heavy neutrino. In Ref. \cite{veg13}, for the case of beta decay, a ratio was considered between the contributions of the heavy and the light neutrino to the differential rates, as we shall see in Section \ref{results_beta}. For the case of electron capture the differential decay rates contain many peaks and it is more convenient to consider integrated decay rates over specific energy ranges, as discussed in \cite{fil14}. In either case (beta decay and electron capture), this amounts to consider that the detector collects events within energy ranges $E_i\pm \Delta$ and $E_j\pm \Delta$ in two regions of the spectrum, such that the mass of the hypothetical heavy neutrino lies in between, namely $E_i+\Delta < m_h < E_j-\Delta$, with $E_j < Q-\Delta$.
A ratio $R$ between the number of collected events in both regions is then performed, which corresponds to the theoretical expression:
\begin{equation}
R = \frac{\Lambda_{i}}{\Lambda_{j}} = \frac{\kappa_i^l + \tan^2\zeta \:\kappa_i^h}{\kappa_j^l} \:,
\label{ratio1}
\end{equation}
where $\Lambda_{i} = \cos^2\zeta \:\kappa^l_{i} + \sin^2\zeta \:\kappa^h_{i}$ (in the region where both $l$ and $h$ mass eigenstates contribute) and $\Lambda_{j} = \cos^2\zeta \:\kappa^l_{j}$ (in the region where $h$ is not energetically allowed). The integrals $\kappa$ are defined as:
\begin{eqnarray}
\kappa^{\nu}_{i,j} = \int_{E_{i,j}-\Delta}^{E_{i,j}+\Delta} \:\frac{d\lambda}{dE_c}(m_{\nu}) \:dE_c
\label{def_kappa}
\end{eqnarray}
In electron capture the energies $i$ and $j$ can be selected as the energies of two peaks and the integration intervals can be chosen as the width of the capture peaks. If the peaks are approximated by delta functions, the integrals (using $E_H=E_i$ and $\Delta\to 0$) can be written as:
\begin{eqnarray}
\kappa^{\nu\:EC}_{i;r} = K_{EC} \:C_{i;r} \:S^{(\nu)}_{i;r} \:(Q_{r}-E_{i})^2 \left[ 1- \left( \frac{m_{\nu}}{Q_{r}-E_{i}}\right)^2 \right]^{1/2} \;,
\label{kappa_capture}
\end{eqnarray}
where $Q_r$ is the atomic mass difference in a given isotope $r$, $E_i$ is the energy position of a given peak $i$ in the calorimeter spectrum, $m_{\nu}$ is $m_{l}\approx 0$, the mass of a light neutrino, or $m_{h}$, the mass of a heavy neutrino, and $S^{(\nu)}_{i;r}$ contains the neutrino momentum dependence for the peak $i$ coming from the leptonic matrix element squared. As explained in the previous section, for allowed transitions, $S^{(\nu)}_{i;r}=1$. For first forbidden transitions some of the capture peaks have $S^{(\nu)}_{i;r}=p^2_{\nu\:i;r}$ for the $i$ peaks corresponding to $s_{1/2}$ shells (as well as for those corresponding to $p_{1/2}$ shells that contribute through their admixtures with the $l=0$ orbital due to relativistic corrections). The other peaks have $S^{(\nu)}_{i;r}=1$ where $i$ corresponds to $p_{3/2}$ shells (and $d_{3/2}$ shells through their admixtures with the $l=1$ orbital due to relativistic corrections).

For electron capture, it is convenient to define a ratio similar to the one in Eq. (\ref{ratio1}) but where the numerator and the denominator are themselves ratios of numbers of events within the same peak but for different isotopes of the same element, $r$ and $s$, so that some atomic corrections cancel out \cite{fil14}:
\begin{equation}
R' = \frac{\Lambda_{i;r}/\Lambda_{i;s}}{\Lambda_{j;r}/\Lambda_{j;s}} = \left(R^l_{ij;rs}\right)^{2(\gamma+1)}  \:\left(\frac{1 + \:\omega_{i;r}^{2\gamma +1} \:\tan^2\zeta}{1 + \:\omega_{i;s}^{2\gamma +1}\:\tan^2\zeta} \right)\:,
\label{ratio2}
\end{equation}
where one should notice that the factors $C_i$ and $C_j$ cancel out in addition to the factor $K_{EC}$ which cancels out in both ratios ($R$ and $R'$). The factors in Eq. (\ref{ratio2}) have very simple analytical forms when one uses Eq. (\ref{kappa_capture}):
\begin{equation}
R^l_{ij;rs} = \frac{(Q_{r}-E_i) (Q_{s}-E_j)}{(Q_{r}-E_j) (Q_{s}-E_i)} \qquad \text{and} \qquad \omega_{i;r(s)} = \left[ 1- \left( \frac{m_h}{Q_{r(s)}-E_i}\right)^2 \right]^{1/2} \;,
\label{Rl_and_w}
\end{equation}
where $\gamma$ depends on the angular momenta of each lepton in a given $\Delta J^{\pi}$ nuclear transition. For instance, for allowed decays, $\gamma=0$, whereas for first forbidden decays, $\gamma=1$ for $s_{1/2}$ and $p_{1/2}$ peaks or $\gamma=0$ for $p_{3/2}$ and $d_{3/2}$ peaks. In the expressions above we have assumed that both peaks $i$ and $j$ are of the same type, namely $\gamma_i=\gamma_j=\gamma$.
 
By measuring the ratio in Eq. (\ref{ratio2}) the mixing angle can then be obtained as
\begin{equation}
\zeta = \arctan \left[\frac{\left(R^l_{ij;rs}\right)^{2(\gamma+1)} - R'_{exp}}{R'_{exp} \:\omega^{2\gamma+1}_{i;s} - \left(R^l_{ij;rs}\right)^{2(\gamma+1)} \:\omega^{2\gamma+1}_{i;r}} \right]^{1/2} \;,
\label{zeta}
\end{equation}
where the theoretical values of $R^l_{ij;rs}$, $\omega_{i;r}$ and $\omega_{i;s}$ require an accurate experimental knowledge of the position of the selected peaks and, more importantly, of the atomic mass differences $Q_r$ and $Q_s$. In addition, one has to consider fixed, initially unknown, values of the mass of the heavy neutrino $m_h$ that is searched for, or otherwise be content with the generation of exclusion plots in case of no effect observation.

In summary, we consider here two types of ratios that can be useful in the search for a signal of keV sterile neutrinos in electron capture: 1) The ratio between the intensity of two peaks in the electron capture spectrum of a given nucleus, as for the case of $^{163}$Ho discussed in Section \ref{results_ec}. 2) The case in which two isotopes of a given element undergo electron capture where one can use the ratio defined in Eq. (\ref{ratio2}), as is the case of Lead isotopes discussed in Section \ref{results_ec}. 
For the beta decay case, in Section \ref{results_beta} we consider the ratio between the differential rates, but the ratios between integrated rates are also useful and may be more realistic for comparison to experiments.

\section{Results for electron capture spectra \label{results_ec}}

Let us first consider the capture of an atomic electron by the nucleus Holmium 163 ($^{163}$Ho; $Z = 67$, $N = 96$) to turn into Dysprosium 163 ($^{163}$Dy; $Z = 66$, $N = 97$), with ground state spins and parities $J^{\pi}=7/2^{-}$ and $J^{\pi}=5/2^{-}$, respectively. It is an allowed unique Gamow-Teller transition ($\Delta J =1$ and no parity change, the leptons carry no orbital angular momentum, $\Delta L=0$, and couple to spin $S=1$), with atomic mass difference $Q=$2.833 $\pm$ 0.030 $\pm$ 0.015 keV \cite{eli15}, that is being currently measured at the Electron Capture Holmium (ECHo) experiment \cite{echo}. Only electrons with principal quantum number $n$ larger than 2 have a binding energy lower than the $Q$-value and can therefore be captured. In addition, being an allowed decay, only electrons from $s_{1/2}$ and $p_{1/2}$ orbitals can be captured, the latter through admixtures with the $l=$ 0 orbital due to relativistic corrections. Thus, capture from the orbitals M1, M2, N1, N2, O1, O2, P1 (in spectroscopic notation) is possible \cite{ruj81,lus11,rob15,fae15}.

\begin{figure}
\centering
\includegraphics[width=0.6\textwidth,angle=0]{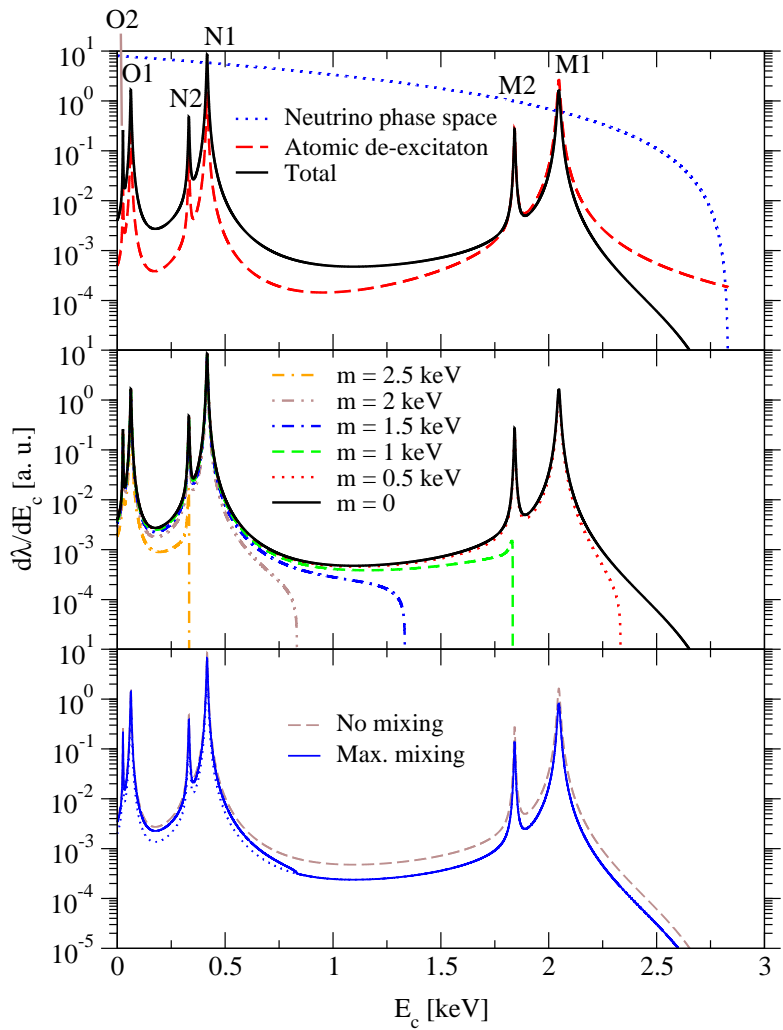}
\caption{Calorimeter spectrum after electron capture in $^{163}$Ho. Upper panel: full spectrum (solid curve) and contributions: emitted neutrino phase space (dotted curve) and daughter atom de-excitation (dashed curve). Middle panel: full spectrum for different neutrino masses: 0, 0.5, 1, 1.5, 2 and 2.5 keV. Lower panel: full spectrum for light ($m_l\approx$ 0 keV) - heavy ($m_h=$ 2 keV) neutrino mixing using a maximal mixing angle $\zeta=45^{\circ}$ for illustration, with (solid curve) and without (dotted curve) the heavy neutrino contribution. For comparison, the spectrum without heavy neutrino mixing (dashed curve) is also shown.
We use the maximal unrealistic value $\zeta=45^{\circ}$ to make the effect visible in the scale of the figure. Realistic values of $\zeta < 0.01^{\circ}$ result in a reduction of the difference between the dashed and the solid lines in the third panel by more than seven orders of magnitude.}
\label{holmio163}
\end{figure}

In Fig. \ref{holmio163} we show the differential electron capture rate in $^{163}$Ho as a function of the calorimeter energy, $E_c=Q-E_{\nu}$. In Table \ref{ho} we give theoretical values of the strengths and widths of each capture peak and experimental values of the electron binding energy again for each capture peak. The strengths $C_H$ are obtained in \cite{toi,bam77} from relativistic Hartree-Fock calculations of the electron wave functions at the nuclear site, accounting for exchange and overlap contributions. In the upper plot the emitted neutrino phase space contribution and the daughter atom de-excitation contribution are plotted separately, together with the product of both, which is the full differential electron capture rate. In the middle plot we show the full differential rate for different masses of the emitted neutrino: $m_{\nu} \approx$ 0 (realistic) and $m_{\nu} \approx$ 0.5, 1, 1.5, 2, and 2.5 keV (unrealistic). As  can be seen in the plot, each curve ends at $Q-m_{\nu}$. Finally, the lower plot shows the spectrum resulting from the mixing of a light neutrino mass eigenstate, $m_{l} \approx$ 0, with a heavy mass eigenstate $m_{h} \approx$ 2 keV (solid curve) as in Eq. (\ref{spectrum_mixing}). Although the spectrum with mixing is realistic in the sense that it is the result expected if a heavy mass eigenstate exists, it is unrealistic in the degree of mixing shown in the figure, that has been maximized here for the sake of visibility of the `kink' in the scale of the figure: 50$\%$ light neutrino and 50$\%$ heavy neutrino, corresponding to a mixing angle $\zeta=$ 45$^{\circ}$. The `kink' can be seen in this curve at $E_c=Q-m_h=$ 0.833 keV. Above this calorimeter energy the heavy mass eigenstate cannot be produced due to energy conservation. For comparison, the spectrum for $m_{\nu}\approx$ 0 with no mixing with heavy states is shown in the dashed curve. 
\begin{table}
\begin{center}
\caption{Atomic parameters for electron capture in Holmium from shells M1 to O2. The values of $C_H$ \cite{toi,bam77} correspond to the atomic shells in Holmium, while the orbital electron widths $\Gamma_H$ \cite{cam01} and binding energies $B_H$ ($< Q$) \cite{bind} correspond to the atomic shells in Dysprosium.}
\label{ho}
\begin{tabular}{c||ccccccc}
 &$\quad$ M1 $\quad$ & $\quad$ M2 $\quad$ & $\quad$ N1$\quad$ &$\quad$ N2 $\quad$ &$\quad$ O1 $\quad$ & $\quad$ O2 $\quad$  \\
\hline
\hline
$\quad$ $C_H$ $\quad$ & 0.05377 & 0.002605 & 0.01373 & 0.0005891 & 0.001708 & 0.0001 &  \\
$\Gamma_H$ [keV] & 0.013 & 0.006 & 0.006 & 0.005 & 0.005 & 0.002 \\
$B_H$ [keV] & 2.047 & 1.842 & 0.416 & 0.332 & 0.063 & 0.026 \\
\end{tabular}
\end{center}
\end{table}

Another example of electron capture under study here is that of the Lead isotopes 202 ($^{202}$Pb; $Z=82$, $N=120$) and 205 ($^{205}$Pb; $Z=82$, $N=123$) going to the Thallium isotopes 202 ($^{202}$Tl; $Z=81$, $N=121$) and 205 ($^{205}$Tl; $Z=81$, $N=124$), respectively, where the process is in both cases a first forbidden Gamow-Teller unique transition, with $0^+ \to 2^-$ for $^{202}$Pb and $5/2^- \to 1/2^+$ for $^{205}$Pb ($\Delta J=2$ and parity change, the leptons carry $\Delta L=1$ and couple to $S=1$). The atomic mass differences obtained from the data in \cite{audi12} are $Q=46\pm 14$ keV and $Q=50.6\pm 1.8$ keV, respectively. According to the explanations given in Section \ref{capture}, captures from orbitals L1, L2, M1, M2, N1, N2, O1, O2 contain an extra neutrino momentum dependence $S^{(\nu)}=p^2_{\nu}$ that modifies the calorimeter spectrum, and correspond to $\gamma=1$. On the other hand, captures from orbitals L3, M3, M4, N3, N4, O3, O4, that correspond to $\gamma=0$, contain an extra electron momentum dependence $p^2_{e}$ which, being fixed in bound electrons, is included in the values of $C_H$. They are given in Table \ref{capture_lead} together with the widths and the experimental electron binding energies in the daughter atom, Thallium.

\begin{figure}
\centering
\includegraphics[width=0.6\textwidth,angle=0]{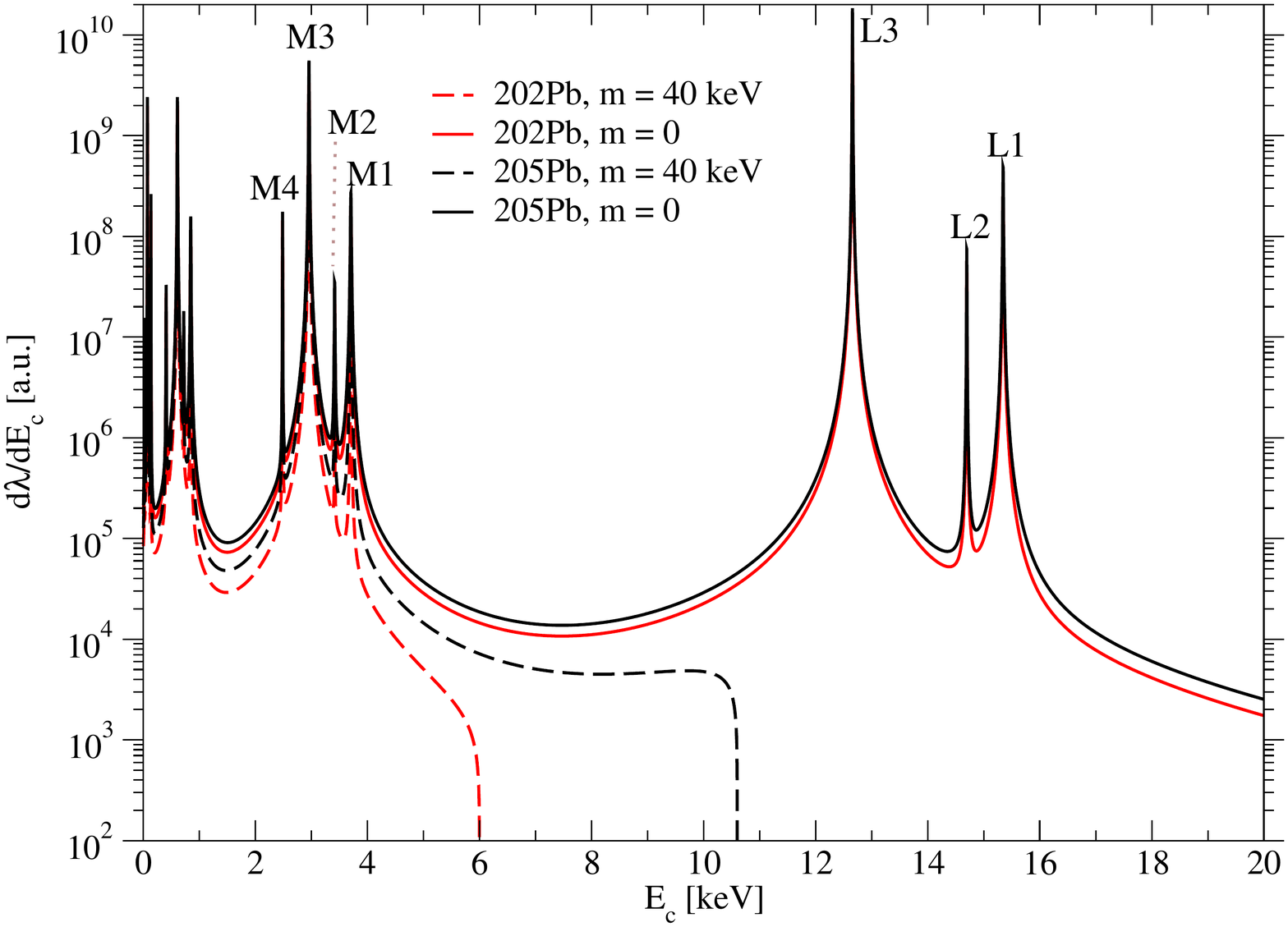}
\\
\includegraphics[width=0.6\textwidth,angle=0]{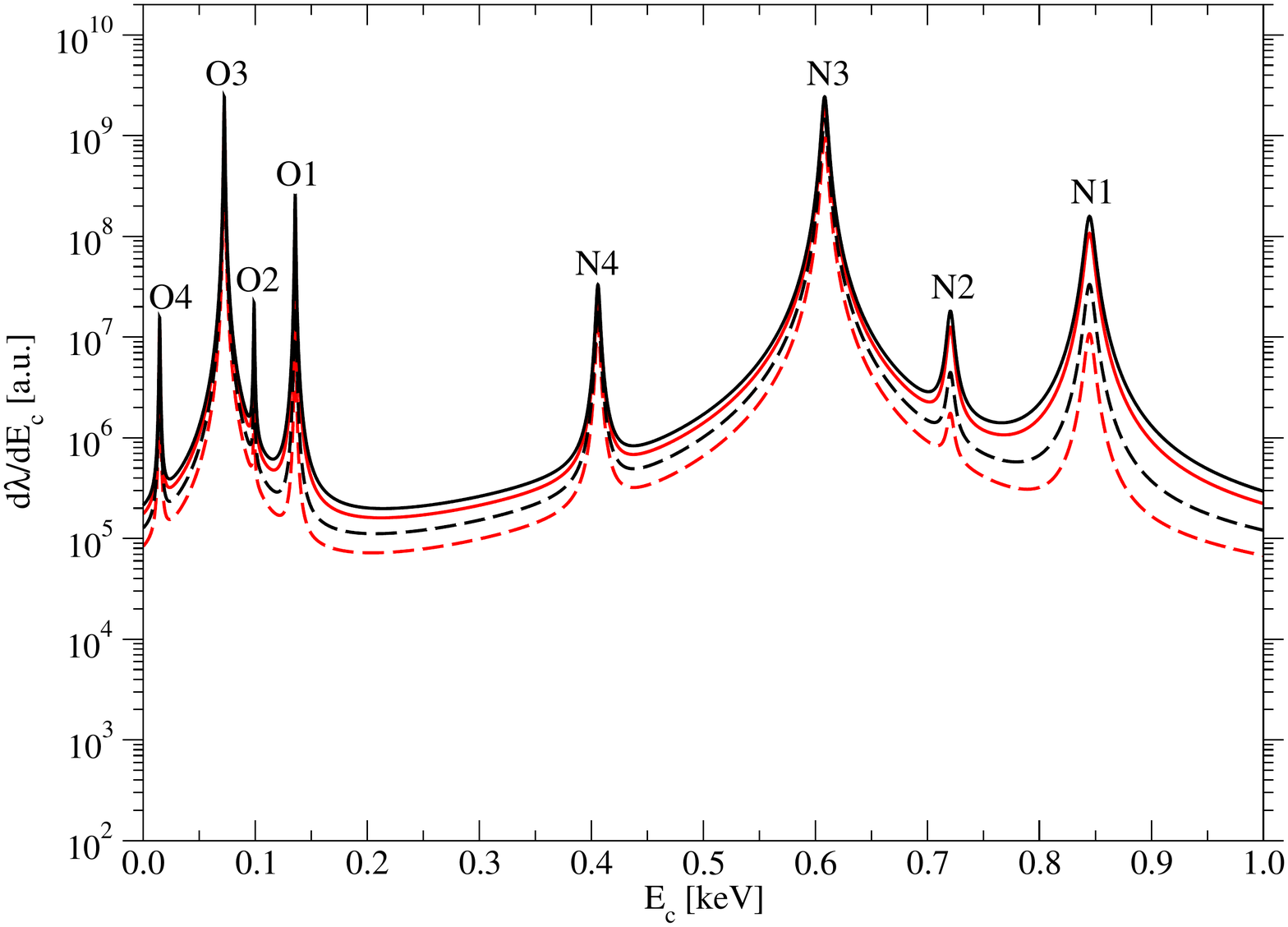}
\caption{Calorimeter spectrum after electron capture in $^{202}$Pb (light curves) and in $^{205}$Pb (black curves) with emission of a light neutrino, $m_{l}=0$ (solid curves) and a heavy neutrino, $m_{h}=40$ keV (dashed curves). Upper plot: full spectrum, showing L and M capture peak labels. Lower plot: low energy region ($E_c=0-1$ keV), showing N and O capture peak labels.}
\label{plomo}
\end{figure}

Fig. \ref{plomo} shows the capture differential rate in $^{205}$Pb (dark curves) and in $^{202}$Pb (light curves). Position, width and strength of the capture peaks are assumed to be the same in both isotopes, but the spectra are different because of the different $Q$-values. Solid curves are for light neutrino emission, $m_{\nu} \approx$ 0, and dashed lines are for heavy neutrino emission with $m_{\nu} =$ 40 keV (results are given separately for each mass, before mixing). 
Due to the large $Q$-values, electron capture in Lead isotopes allows us to explore this mass value, although it may be somewhat larger than the expected value from cosmological reasons.
The analysis of the ratio in Eq. (\ref{ratio2}) can be computed, for example, using the capture peaks L3 at $E_c=$ 12.657 keV and M3 at $E_c=$ 2.956 keV, both being of the $\gamma=0$ type. For $m_h=$ 40 keV one would obtain $R^l_{M3\:L3; \:202\:205}=1.037$, $\omega_{M3;\:202}=0.369$, $\omega_{M3;\:205}=$ 0.543. These results should be introduced in Eq. (\ref{zeta}), together with the experimental ratio $R'_{exp}$, to obtain the value of the mixing angle $\zeta$.

\begin{table}
\begin{center}
\caption{Same as Table \ref{ho} (data from the same references) but for electron capture in Lead going to Thallium. The upper table summarizes the values of the relevant parameters for $s_{1/2}$ and $p_{1/2}$ atomic shells, while the lower table shows the values for $p_{3/2}$ and $d_{3/2}$ atomic shells. In the latter case, $C_H$ includes an extra factor $p^2_e$.}
\label{capture_lead}
\begin{tabular}{c||cccccccc}
 &$\quad$ L1 $\quad$ &$\quad$ L2 $\quad$ &$\quad$ M1 $\quad$ & $\quad$ M2 $\quad$ & $\quad$ N1$\quad$ &$\quad$ N2 $\quad$ &$\quad$ O1 $\quad$ & $\quad$ O2 $\quad$ \\
\hline
\hline
$\quad$ $C_H$$\quad$ & 0.8781 & 0.06647 & 0.2125 & 0.01759 & 0.05849 & 0.004431 & 0.01016 & 0.0008052 \\
$\Gamma_H$ [keV] & 0.011 & 0.006 & 0.015 & 0.010 & 0.009 & 0.007 & -- & -- \\
$E_H$ [keV] & 15.346 & 14.697 & 3.703 & 3.415 & 0.845 & 0.720 & 0.136 & 0.099   \\
\\
 &$\quad$ L3 $\quad$ &$\quad$ M3 $\quad$ &$\quad$ M4 $\quad$ & $\quad$ N3 $\quad$ & $\quad$ N4 $\quad$ &$\quad$ O3 $\quad$ &$\quad$ O4 $\quad$\\
\hline
\hline
$\quad$ $C_H$ [keV$^2$] $\quad$ & 18978.193 & 5366.014 & 45.383 & 1363.307 & 12.607 & 237.280 & 1.500 \\
$\Gamma_H$ [keV] & 0.006 & 0.009 & 0.002 & 0.006 & 0.004 & 0.001 & 0.001 \\
$B_H$ [keV] & 12.657 & 2.956 & 2.484 & 0.608 & 0.406 & 0.072 & 0.015   \\
\end{tabular}
\end{center}
\end{table}

It is important to remark that the theoretical quantities $R^l$ and $\omega$ computed as described above contain several approximations. They correspond to Dirac-delta peaks, ignoring the actual shapes and widths, and thus they just refer to one single capture peak, neglecting the possible effect of the tails of nearby one-hole peaks. Moreover, the influence of two-hole peaks close to the main ones has also been neglected. Other effects that have not been taken into account but whose influence is expected to be very small are multi-hole (more than two) peaks, virtual intermediate states (influence of transitions through atomic shells not energetically accesible) or interference between atomic transitions resulting from addition of amplitudes instead of intensities \cite{ruj13}. The accurate experimental determination of the ratio $R'_{exp}$ entails its own difficulties, among them the possible existence of metastable atomic states whose delayed de-excitations fail to contribute to the collected spectrum, and the variety of chemical environments resulting in a complex mixture of $Q$-values.

\section{Results for beta decay spectra \label{results_beta}}

In addition to electron capture, we also show here the effect of heavy neutrino emission in the electron spectrum of beta decays for one of the cases of most interest, Tritium. It has been studied in depth in previous works \cite{veg13,mer15}, together with another process that has also drawn considerable theoretical end experimental attention, the beta decay of Rhenium 187 \cite{adh16,mar,veg13}.
The beta decay of Tritium ($^{3}$H; $Z$ = 1, $N$ = 2) going to Helium 3 ($^{3}$He; $Z$ = 2, $N$ = 1) is
\begin{equation}
^{3}\text{H} \:\rightarrow \:^{3}\text{He} + e^- + \bar{\nu}_{e} \:,
\label{tritio}   
\end{equation}
and has a $Q$-value of 18.59 keV \cite{audi12}. Both the initial and the final nuclear ground states have spin-parity $1/2^{+}$, and the transition is allowed with the electron and the antineutrino emitted in $s$-wave. The Karlsruhe Tritium Neutrino Experiment (KATRIN) \cite{kat} is currently studying this decay in order to determine the active neutrino mass, and could also study the production of a heavy neutrino of a mass lower than 18 keV.

The differential decay rate with respect to the electron energy is given by
\begin{equation}
\frac{d\lambda}{dE_e} =  K_{\beta}  \:(E_e^2-m_e^2)^{1/2} \:\left[(Q+m_e-E_e)^2-m_{\nu}^2 \right]^{1/2} \:E_{e} \:(Q+m_e - E_{e})
\label{spectrum_beta}   
\end{equation}
where $K_{\beta}$ contains, among others, the weak interaction coupling, the nuclear matrix element and the Fermi function.

\begin{figure}[h]
\centering
\begin{minipage}{1.0\textwidth}
\centering
\includegraphics[width=0.49\linewidth]{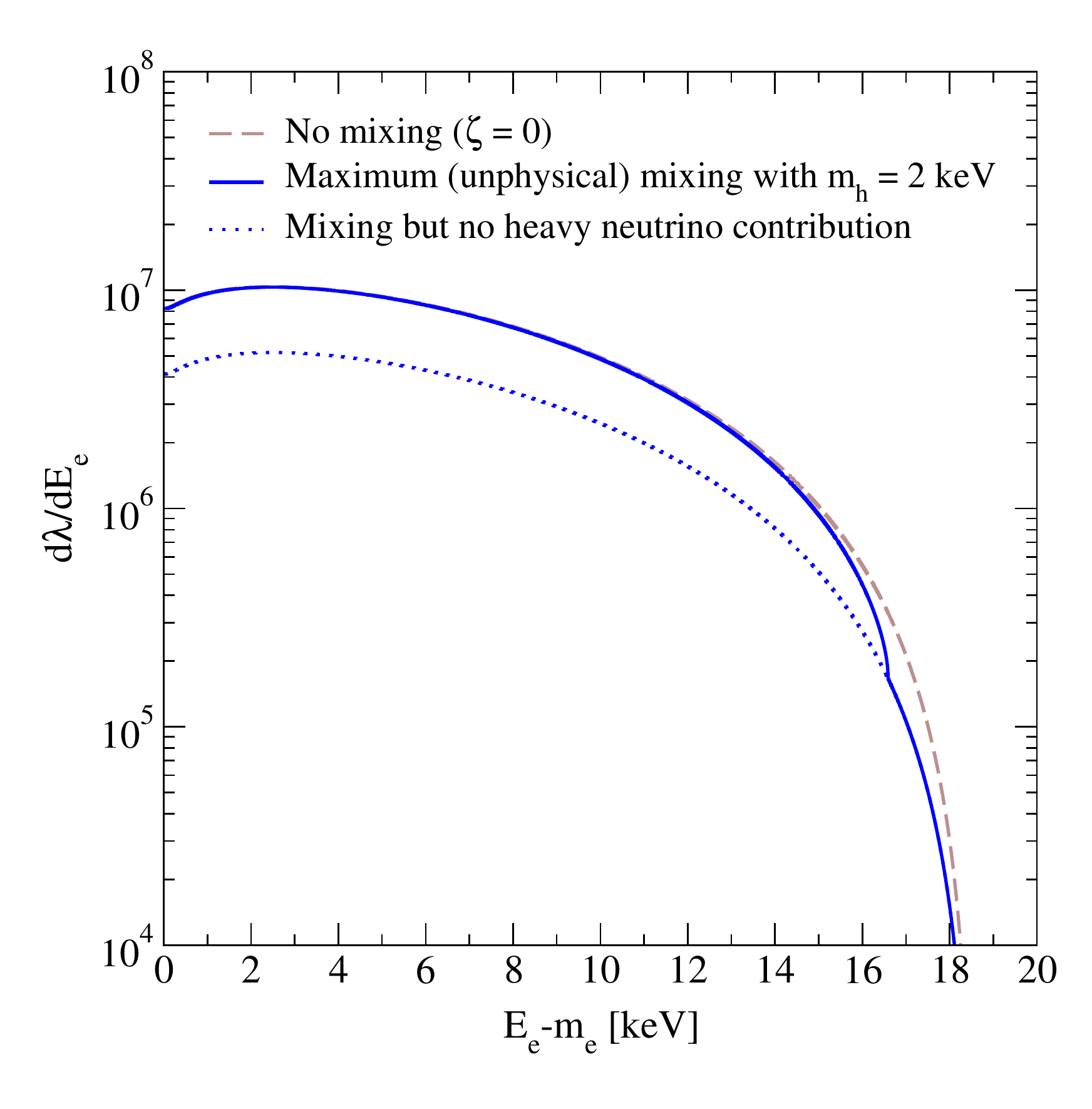}
\includegraphics[width=0.49\linewidth]{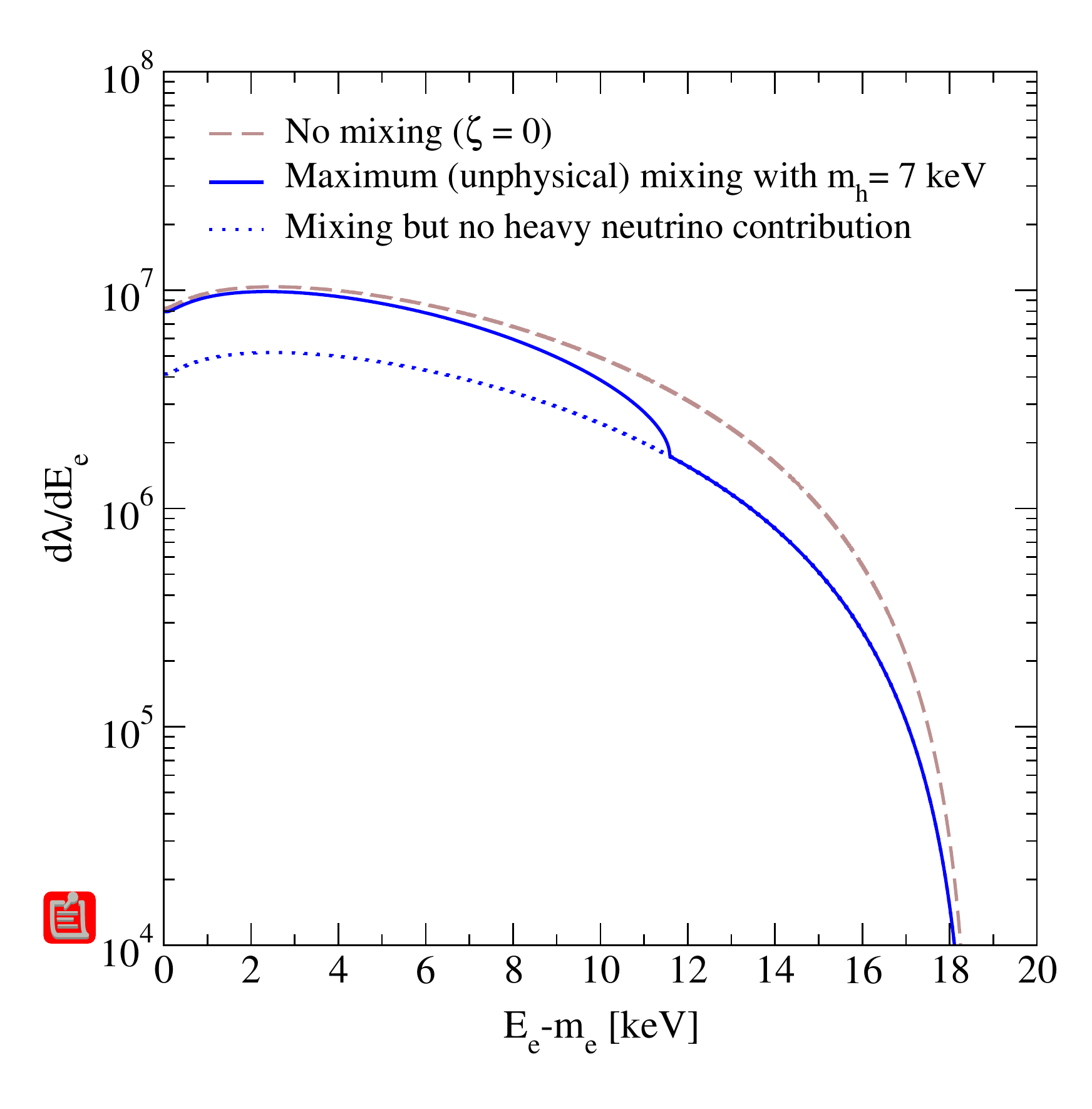}
\caption{Electron spectrum of Tritium beta decay for a light-heavy neutrino mixing angle $\zeta = 45^{\circ}$, shown just for illustration, and a heavy neutrino mass (solid curve) $m_h=$ 2 keV (left plot) and $m_h=$ 7 keV (right plot). The spectrum without heavy neutrino mass contribution (dotted curve) and without mixing (dashed curve) are also shown.
We use the maximal unrealistic value $\zeta=45^{\circ}$ to make the effect visible in the scale of the figure. Realistic values of $\zeta < 0.01^{\circ}$ result in a reduction of the difference between the dotted and the solid lines by more than seven orders of magnitude.}
\label{sterile_Gmixing_tritio}
\end{minipage}
\end{figure}

As an illustration of the heavy neutrino effect in the Tritium beta decay, we plot in Fig.~\ref{sterile_Gmixing_tritio} the differential decay rate from Eqs. (\ref{spectrum_beta}) and (\ref{spectrum_mixing}) using a maximal mixing with an unrealistic value of the mixing angle $\zeta = 45^{\circ}$, to show the effect more distinctly in the scale of the figure. We have used heavy mass components with $m_h = 2$ keV \cite{men16} (plot on the left) and with $m_h = 7$ keV \cite{aba14} (plot on the right). For comparison, the spectra without the heavy neutrino contribution ($\lambda^h = 0$) and without mixing ($\zeta = 0^{\circ}$) are also shown in this plot. The kink in the spectrum at $ E_e-m_e = Q-m_h$ can be observed if the experimental relative error is lower than the size of the step, the latter being very small in realistic situations.

The effect of a heavy neutrino emission can be analyzed through the ratio between the heavy and the light neutrino contributions to the spectrum:
\begin{equation}
{\mathcal R} \equiv \:\frac{d\lambda^{h}/dE_{e}}{d\lambda^{l}/dE_{e}}  \tan^2\zeta 
\label{ratio_beta}
\end{equation}
The full differential decay rate is also related to the ratio ${\mathcal R}$ through
\begin{equation}
\frac{d\lambda}{dE_e} = \frac{d\lambda^{l}}{dE_e}  \:[1 +  {\mathcal R} ]\:  \cos^2\zeta
 \label{ratio_beta2} \; ,
\end{equation}

\begin{figure}
\centering
\includegraphics[width=0.6\textwidth,angle=0]{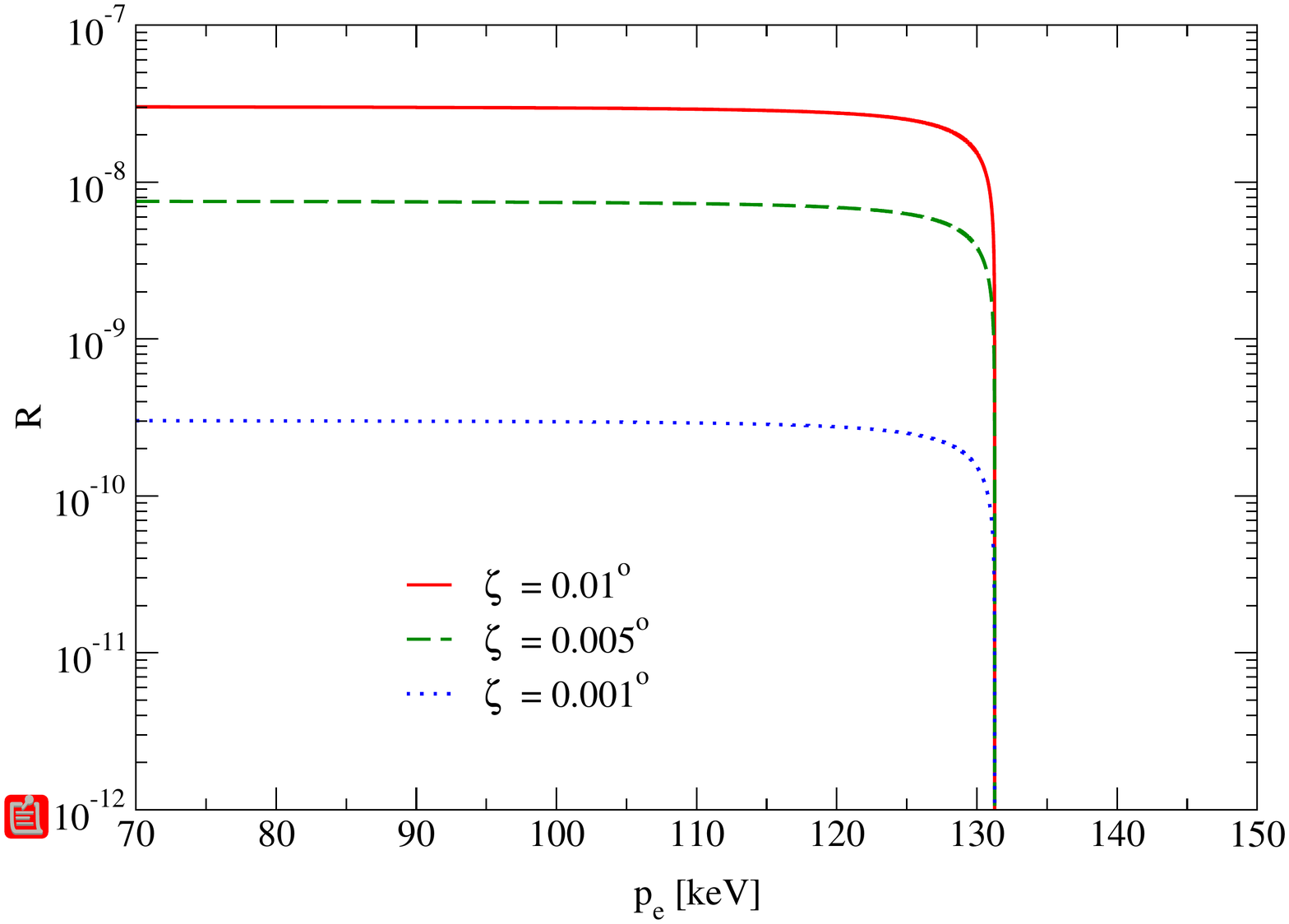}
\caption{Ratio ${\mathcal R}$, defined in Eq. (\ref{ratio_beta}), for the Tritium beta decay, as a function of the electron momentum for a heavy neutrino mass $m_h=$ 2 keV and different values of the mixing angle: $\zeta=$ 0.01$^{\circ}$ (solid line), 0.005$^{\circ}$ (dashed line) and 0.001$^{\circ}$ (dotted line).}
\label{sterile_R_m2kev_tritio}
\end{figure}

In Fig.~\ref{sterile_R_m2kev_tritio} we plot ${\mathcal R}$, the ratio of the heavy neutrino contribution over the light neutrino contribution to the decay rate, as a function of the momentum of the emitted electron for a heavy neutrino mass $m_h = 2$ keV and different mixing angles: $\zeta=$ 0.01$^{\circ}$, 0.005$^{\circ}$ and 0.001$^{\circ}$. The size of this ratio ($< 10^{-7}$) gives an idea of the difficulty of finding the kink in the spectrum due to the production of the heavy mass eigenstate. As can be seen in the figure, the ratio is different from zero and almost constant in the range $0 \leq p_e < (p_e)_{\text{max}}$, where the maximum electron momentum is given by $(p_e)_{\text{max}} = [(Q-m_{h}) (Q-m_{h} + 2 m_e)]^{1/2}$.  For the heavy mass used in the figure, $(p_e)_{max} \simeq$ 131.3 keV. The ratio ${\mathcal R}$ decreases as the mixing angle decreases, being approximately proportional to $\zeta^2$. It also decreases for larger heavy neutrino masses $m_h$. A different ratio between the spectra with mixing and without mixing (${\mathcal R}^*$) has also been considered in \cite{veg13,mer15}, related with ${\mathcal R}$ in Eq. (\ref{ratio_beta}) as
\begin{equation}
{\mathcal R}^* = -\sin^2 \zeta + {\mathcal R} \:\cos^2 \zeta = \frac{d\lambda/dE_e}{d\lambda^l/dE_e}  - 1
\label{ratio_star}
\end{equation}

\section{Conclusions \label{conclusions}}
Signatures of hypothetical keV sterile neutrinos, which could be warm dark matter candidates (WDM), may be found in the spectra of ordinary weak nuclear decays. The electron spectrum in beta decay is cleaner for this purpose because it is a smooth curve, while the de-excitation spectrum after electron capture shows many peaks. On the other hand in beta decay the possible excitation of the atoms or molecules are not disentangled while in electron capture the calorimeter collects all the energy independently of the excitation of the atom and of the de-excitation path, and one has higher statistics around the capture peaks.

Figures of electron spectrum for beta decay in $^3$H as well as of calorimeter spectrum for electron capture in $^{163}$Ho and $^{202}$Pb and $^{205}$Pb are given considering various values of the heavy neutrino mass (2 keV, 7 keV and 40 keV) that may be experimentally probed. In both weak processes the small value of the light-heavy neutrino mixing angle requires extremely high experimental precision and the use of sources with large stability to reduce systematic errors. This is why it is useful to consider relevant ratios between transition rates as first introduced in \cite{fil14}, \cite{veg13} to analyze the data. We consider two cases: the case of a single isotope, where one may use the ratio of accumulated number of events in different regions of the spectrum, which allows us to remove uncertainties related to the nuclear matrix element and to the values of overall constants. The case of two isotopes, where we consider ratios of the above mentioned ratios that allow us to reduce uncertainties from atomic parameters. We give as well analytical expressions that can be used to obtain a good theoretical approximation to the experimental ratios (Eqs. (\ref{kappa_capture}) to (\ref{Rl_and_w})). 

Both the experimental measurement and the theoretical model must be accurate enough to detect differences in the expected versus the measured ratios of the order of the mixing angle squared, $\zeta^2 \lesssim 10^{-8}$. This is also the size of the kink expected in the electron spectrum of beta decay, located at the limit of the region where the production of a heavy neutrino is energetically allowed. In order to identify this kink among the statistical fluctuations of the measured spectrum, the number of collected events must be larger than the inverse of the squared ratio ${\mathcal R}$ in Eq. (\ref{ratio_beta}).

Our results, particularly on the electron capture spectra of Lead isotopes that include for the first time all the possible peaks, may help in designing and analyzing future experiments to search for sterile neutrinos in the keV mass range. The use of the different types of ratios between number of events discussed here, for both electron capture and beta decay processes, may also help in planning the experiments by establishing the threshold of statistical and systematic uncertainties, required for the detection of sterile neutrinos of given mass and mixing, or to properly extract exclusion plots.

\acknowledgments

E.M.G. and O.M. acknowledge MINECO for partial financial support (FIS2014-51971-P) and IEM-CSIC for hospitality. O.M. acknowledges support from a Marie Curie International Outgoing Fellowship within the European Union Seventh Framework Programme, under Grant Agreement PIOF-GA-2011-298364 (ELECTROWEAK).


\begin{thebibliography}{}

\bibitem{zwi33}
F. Zwicky, Physica Acta 6 (1933) 124; M. Roos, arXiv:1208.3662 [astro-ph].

\bibitem{cosmo}
S. Dodelson, {\it Modern Cosmology}, Academic Press (2003); S. Weinberg, {\it Cosmology}, Oxford University Press (2008).

\bibitem{les12}
J. Lesgourgues and S. Pastor, Adv. High Energy Phys. 2012 (2012) 608515.

\bibitem {boy08}
D. Boyanovsky, H. J. de Vega, N. G. S\'anchez, Phys. Rev. D 77 (2008) 043518; Phys. Rev. D 78 (2008) 063546; C. Destri, H. J. de Vega, N. G. S\'anchez, Phys. Rev. D 88 (2013) 083512; H. J. de Vega, P. Salucci, N. G. S\'anchez, New Astronomy 17 (2012) 653; Mon. Not. R. Astron. Soc. 442 (2014) 2717; H. J. de Vega, N. G. S\'anchez, Mon. Not. R. Astron. Soc 404 (2010) 885;  Int. J. Mod. Phys. A 26 (2011) 1057; Phys. Rev. D 85 (2012) 043516, 043517; Int  Jour. Mod. Phys. A 31 (2016) 1650073; N. Menci, S. Paduroiu, P. Salucci, N. S\'anchez, C. R. Watson, lectures at the 20th Paris Cosmology Colloquium Chalonge - H\'ector de Vega 2016. http://chalonge.obspm.fr/

\bibitem{kolb90}
E. W. Kolb, M.S. Turner, {\it The Early universe}, Addison Ð Wesley (1990).

\bibitem{cusped}
J. F. Navarro, C.S. Frenk, S.D. White, Astrophysical Journal 940 (1997) 493.

\bibitem{cored}
W. J. G. de Blok, Advances in Astronomy 2010, 1; arXiv:09103538;
G. Gilmore {\it et al.}, Astrophysical Journal 663 (2007) 948;
P. Salucci, Ch. Frigerio Martins, EAS Publication Series 36 (2009) 133;
J. van Eymeren {\it et al.}, A \& A 505 (2009) 1;
M. Walker, J. Pe\~narrubia, Astrophysical Journal 742 (2011) 20;
R. F. G. Wyse, G. Gilmore, IAU Symposium 244 (2007) 44;
A. Burkert, Astrophysical Journal 447 (1995) 25.
 
\bibitem{men16}
N. Menci, N.G. S\'anchez, M. Castellano, A. Grazian, Astrophysical Journal  818 (2016) 90.

\bibitem{bulboy14}
E. Bulbul, M. Markevitch, A. Foster, R. K. Smith, M. Loewenstein and S. W. Randall, Astrophysical Journal 789 (2014) 13; A. Boyarsky, O. Ruchayskiy, D. Iakubovskyi and J. Franse, Phys. Rev. Lett. 113 (2014) 251301.

\bibitem{aba14}
K. N. Abazajian, Phys. Rev. Lett. 112 (2014) 161303; N. Menci, A. Grazian, M. Castellano and N. G. S\'anchez, Astrophysical Journal Letters 825 (2016) 1.

\bibitem{des13}
C. Destri, H. de Vega, N. G. S\'anchez, Astroparticle Physics 46 (2013) 14.

\bibitem {dol02}
A. D. Dolgov, Phys. Rep. 370 (2002) 333; A. Kusenko, Phys. Rep. 481 (2009) 1.

\bibitem {pon68}
B. Pontecorvo, Sov. Phys. JETP 26 (1968) 984.

\bibitem{mer13}
A. Merle, Int. J. Mod. Phys. D 22 (2013) 1330020.

\bibitem {lsp}
J.D. Vergados and T.S. Kosmas, Phys. Atom. Nucl. 61 (1998) 1066; E. Holmlund, M. Kortelainen, T.S. Kosmas, J. Suhonen and J. Toivanen, Phys. Lett. B 584 (2004) 31.

\bibitem{adh16}
R. Adhikari {\it et al.}, A White Paper on the keV Sterile Neutrino Dark Matter, arXiV: 1602.04816 [hep-ph]  2016.

\bibitem{drex13}
G. Drexlin, V. Hannen, S. Mertens, C. Weinheimer, Adv. High Energy Phys. 2013 (2013) 293986.

\bibitem{mar} MARE collaboration,
http://mare.dfm.uninsubria.it/frontend/exec.php; A. Nucciotti, Neutrino 2010, arXiv:1012.2290. A. Nucciotti, on behalf of the MARE collaboration, lecture at the Workshop Chalonge Meudon 2011. http://chalonge.obspm.fr/

\bibitem{kat} KATRIN collaboration, http://www-ik.fzk.de/tritium/;
C. Weinheimer, Varenna Enrico Fermi Course CLXX,  arXiv:0912.1619; C. Weinheimer,
lecture at the 20th Paris Cosmology Chalonge-H\'ector de Vega Colloquium 2016; G. Drexlin lecture at the 19th Paris Cosmology Colloquium 2015; A. Huber, lecture at the Chalonge-de Vega Meudon Workshop 2016. http://chalonge.obspm.fr/ 

\bibitem{ptolemy}
C. Tully, lecture at the 19th Paris Cosmology Colloquium 2015; S. Betts {\it et al.}, arXiv:1307.4738v2 [astro-ph.IM].

\bibitem{project8}
D. M. Asner {\it et al.}, Phys. Rev. Lett. 114 (2015) 162501.

\bibitem{echo}
P. C.-O. Ranitzsch, J.-P. Porst, S. Kempf {\it et al.}, J. Low Temp. Phys. 167 (2012) 1004; C. Hassel, lecture at the International School of Astrophysics Daniel Chalonge - H\'ector de Vega 2015; L. Gastaldo {\it et al.}, J. Low Temp. Phys. 176 (2014) 876.

\bibitem{holmes}
B. Alpert et al., Eu. Phys. J. C 7 (2015) 112.

\bibitem{fer34}
E. Fermi, Nuov. Cim. 11 (1934) 1; Z. Phys. 88 (1934) 161.

\bibitem{kos15}
Th. Kosmas, H. Ejiri, and A. Hatzikoutelis, Adv. High Energy Phys. 2015 (2015) 806067 and references therein.

\bibitem{giu12}
A. Giuliani and A. Poves, Adv. High Energy Phys. 2012 (2012) 857016.

\bibitem{stellar}
J.-U. Nabi and H. V. Klapdor-Kleingrothaus, Atomic Data and Nuclear Data Tables 71 (1999) 149; Atomic Data and Nuclear Data Tables 88 (2004) 237.

\bibitem{osci}
Y. Fukuda {\it et al.} (Super-Kamiokande collaboration), Phys. Rev. Lett. 81 (1998) 1562; Q.R. Ahmad {\it et al.} (SNO collaboration), Phys. Rev. Lett. 87 (2001) 071301; Q.R. Ahmad {\it et al.} (SNO collaboration), Phys. Rev. Lett. 89 (2002) 011301; G. J. Feldman, J. Hartnell, and T. Kobayashi, Adv. High Energy Phys. 2013 (2013) 475749.

\bibitem{pet13}
S. Petcov, Adv. High Energy Phys. 2013 (2013) 852987.

\bibitem{dk} A. D. Dolgov, Phys. Rep. 370 (2002) 333; A. Kusenko, Phys. Rep. 481 (2009) 1; F. Munyaneza, P. L. Biermann, Astron. and Astrophys. 458 (2006) L9; D. Boyanovsky, C. M. Ho, JHEP 0707 (2007) 030.

\bibitem{shr} R. E. Shrock, Phys. Lett. B 96 (1980) 159; Phys. Rev. D 24 (1981) 1232; AIP
  Conf. Proc. 72 (2008) 368.

\bibitem{con13}
J. M. Conrad, C. M. Ignarra, G. Karagiorgi, M. H. Shaevitz, and J. Spitz, Adv. High Energy Phys. 2013 (2013) 163897.

\bibitem{nuc15}
A. Nucciotti, Adv. High Energy Phys. 2016 (2016) 1.

\bibitem{rob15}
R. G. H. Robertson, Phys. Rev. C 91 (2015) 035504.

\bibitem{ruj13}
A. De R\'ujula, arXiv:1305.4857 [hep-ph]; arXiv: 1601.049902v2 [hep-ph].

\bibitem{fae16}
A. Faessler, F. Simkovic, Phys. Scr. 91 (2016) 043007.

\bibitem{bam77}
W. Bambynek, M. H. Chen, B. Crasemann, M. L. Fitzpatrick, K. W. D. Ledingham, H. Genz, M. Mutterer, R. L. Intemann, Rev. of Mod. Phys. 49 (1977) 77.

\bibitem{fil14}
P. E. Filianin, K. Blaum, S. A. Eliseev, L. Gastaldo, Yu. N. Novikov, V.M. Shabaev, I. I. Tupitsyn, J. Vergados, J. Phys. G: Nucl. Part. Phys. 41 (2014) 095004.

\bibitem{eli15}
S. Eliseev, K. Blaum, M. Block, S. Chenmarev, H. Dorrer, Ch. E. D{\"u}llmann, C. Enss, P. E. Filianin, L. Gastaldo, M. Goncharov, U. K{\"o}ster, F. Lautenschl{\"a}ger, Yu. N. Novikov, A. Rischka, R. X. Sch{\"u}ssler, L. Schweikhard, A. T{\"u}rler, Phys. Rev. Lett. 115 (2015) 062501.

\bibitem{ruj81}
A. De R\'ujula, Nucl. Phys. B 188 (1981) 414.

\bibitem{lus11}
M. Lusignoli, M. Vignati, Phys. Lett. B 697 (2011) 11.

\bibitem{fae15}
A. Faessler, L. Gastaldo, F. Simkovic, J. Phys. G: Nucl. Part. Phys. 42 (2015) 015108.

\bibitem{toi}
R. B. Firestone, V. S. Shirley (Eds.), {\it Table of Isotopes, 8th Ed.}, John Wiley and Sons, Inc. (1999).

\bibitem{cam01}
J. L. Campbell, T. Papp, At. Data Nucl. Data Tables 77 (2001) 1.

\bibitem{bind}
J. A. Bearden, A. F. Burr, Rev. Mod. Phys. 39 (1967) 125; M. Cardona, L. Ley, Eds. {\it Photoemission in Solids I: General Principles}, Springer-Verlag, Berlin, 1978; J. C. Fuggle, N. Martensson, J. Electron Spectrosc. Relat. Phenom. 21 (1980) 275.

\bibitem{audi12}
G. Audi, F. G. Kondev, M. Wang, B. Pfeiffer, X. Sun, J. Blachot and M. MacCormick, Chin. Phys. C 36 (2012) 1157.

\bibitem{veg13}
H.J. de Vega, O. Moreno, E. Moya de Guerra, M. Ram\'on Medrano and N. G. S\'anchez, Nucl. Phys. B 866 (2013) 177; E. Moya de Guerra, O. Moreno, P. Sarriguren and M. Ram\'on Medrano, J. Phys. Conf. Proc. Ser. 366 (2012) 012011; J. M. Boillos, O. Moreno and E. Moya de Guerra, AIP Conf. Proc. 1541 (2013) 167.

\bibitem{mer15}
S. Mertens, JCAP 1502 (2015) 020.

\end{thebibliography}
\end{document}